\documentclass[aps,prb,twocolumn,showpacs,10pt]{revtex4-1}

\usepackage{etoolbox}
\apptocmd{\sloppy}{\hbadness 10000\relax}{}{}
\usepackage{xcolor}
\usepackage{amsmath,amssymb}
\usepackage{graphicx}
\usepackage{subfig}

\newcommand\dg\dagger
\newcommand\R\rangle
\renewcommand\L\langle
\renewcommand\l\left
\renewcommand\r\right

\newcommand\dr{\,\mathrm{d}}

\renewcommand\vector[1]{\mathbf{#1}}

\newcommand\rv{\vector{r}}

\begin{document}

\title{Two-Component Fractional Quantum Hall Effect in the Half-Filled Lowest Landau Level in an Asymmetric Wide Quantum Well}

\author{N. Thiebaut$^1$, M. O. Goerbig$^1$ and N. Regnault$^{2,3}$}

\affiliation{$^1$Laboratoire de Physique des Solides, Universite Paris-Sud, CNRS UMR 8502, 
F-91405 Orsay Cedex, France\\
$^2$Laboratoire Pierre Aigrain, Departement de Physique, ENS, CNRS, 24 rue Lhomond, 
75005 Paris, France \\
$^3$Department of Physics, Princeton University, Princeton, New Jersey 08544, USA}

\date{\today}

\begin{abstract}

We investigate theoretically the
fractional quantum Hall effect at half-filling in the lowest Landau level observed in asymmetric wide quantum wells. 
The asymmetry can be achieved by a potential bias applied between the two sides of the well. Within exact-diagonalization calculations 
in the spherical geometry, we find that the incompressible state is described in terms of a two-component wave function. Its overlap with the ground state
can be optimized with the help of a rotation in the space of the pseudospin, which mimics the lowest two electronic subbands. 

\end{abstract}

\pacs{}

\maketitle


\section{Introduction}

The study of the fractional quantum Hall effect (FQHE) is intimitely bound to progress in the fabrication of
high-mobility samples, and the highest mobilities are achieved today in wide GaAs/AlGaAs quantum wells. 
A drawback of a large well width is, however, the stronger three-dimensional character of the electron motion,
whereas the FQHE is a manifestation of electronic correlations in a strong magnetic field in a \textit{two-dimensional}
(2D) electron system -- strictly speaking, the 2D character is preserved in quantum wells only as long as the electronic-subband separation
is larger than the other relevant energy scales, mainly the Coulomb interaction. In this case, important
insight into the underlying electron liquids and their exotic excitations has
been provided by studying trial wave functions \cite{Laughlin,Jain_CF,Jain,MooreRead} for electrons in the lowest subband 
and in a single spin branch. The latter assumption was soon criticized by Halperin, who pointed out that
the Zeeman effect is insufficient to justify this assumption because it is generally much lower than the leading interaction energy scale. To 
cure this drawback, he proposed \cite{Halperin}
a two-component generalization of Laughlin's wave function 
\cite{Laughlin} that has proven to be extremely fruitful in the understanding of several two-component
quantum Hall systems even if they are of lower symmetry than the original SU(2) spin symmetry. 
As an example beyond spin physics one may mention bilayer 
quantum Hall systems,\cite{Eisenstein,moon} where the layer index can be mimicked by a 
spin 1/2, or more recently graphene with its internal spin-valley degree of freedom.\cite{goerbig_rev}

In order to understand the FQHE in wide quantum wells, while maintaining the 2D aspect of the system, only
very recently a multi-component picture has been utilized, in which the different components consist of the electronic subbands. In the simplest
case, one may restrict the theoretical description to the lowest two subbands, which are then mimicked by a pseudospin $s=1/2$.\cite{Papic} 
Within this model, exact-diagonalization \cite{Papic,Peterson2} and Monte-Carlo calculations \cite{Scarola}
have shed light on a recently observed FQHE at a filling factor $\nu=n_{\rm el}/n_B=1/2$ (and $\nu=1/4$),\cite{Luhman,Shabani1} 
where $n_{\rm el}$ is the electronic and $n_B$ the flux density. The states can be understood
in terms of the so-called (331) [and (553)] Halperin wave function, the precise form of which is 
presented later in this paper [Eq. (\ref{eq:Halperin})]. In addition to the well width, wide quantum wells can be rendered asymmetric
via a potential bias that is experimentally achieved by side gates and that presents a highly controllable parameter in the
experimental study of the FQHE in these systems. Indeed, recent experiments revealed an intriguing transition from compressible to incompressible states
when this potential bias is varied.\cite{Shabani2,Shabani1}

\begin{figure}[!ht]
\includegraphics[scale=0.5]{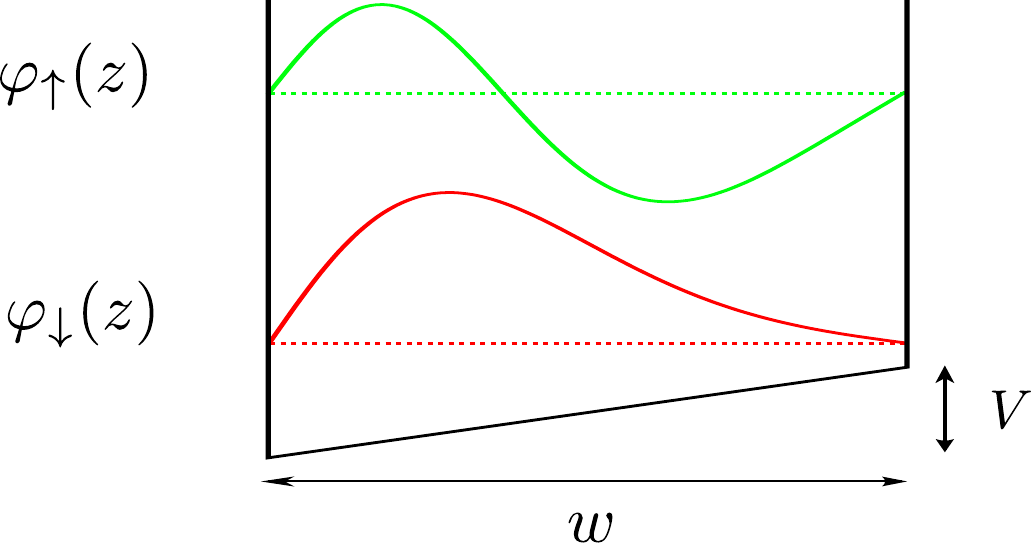}
\caption{(Color online) Scheme of the BQW, the first two eigenfunctions $\varphi_{\uparrow}$ and $\varphi_{\downarrow}$ are plotted. \label{fig:TriangularEigen}}
\end{figure}
Here we investigate the (331) Halperin state within a specially designed model for a biased quantum
well (BQW) that consists of a wide quantum well with infinite potential walls. The potential bias
is modeled by a linear slope $Vz/w$, where $w$ is the well width in the $z$-direction (see Fig. \ref{fig:TriangularEigen}). We perform exact-diagonalization studies of the Coulomb 
interaction within this model in the spherical geometry, where we take into account 
the two lowest electronic levels of the confinement
potential that are mimicked by a pseudospin $s=1/2$. While at this stage the restriction to only two subbands is a model
assumption, we justify it \textit{a posteriori} by an investigation of the pseudospin polarization that indicates a preferential 
occupation of the lowest subband in the relevant parameter range. Our main finding is that the FQHE at $\nu=1/2$ in 
the lowest LL can be interpreted as a (331) Halperin state in a particular region of the $(w,V)$ 
parameter space, surrounded by compressible states.
This result is in excellent agreement with recent experimental findings,\cite{Shabani2,Shabani3} where a FQHE state at 
$\nu=1/2$ was stabilized in an intermediate bias range within a wide quantum well, while it vanishes for very small and large values of $V$. 
On a more technical side, our theoretical approach shows that the overlap of this state with the exact ground state can be largely 
enhanced at a variational parameter that physically plays the role of the pseudospin-polarization angle. 

The paper is organized as follows. We introduce the two-band model of the biased wide quantum in the FQHE regime in Sec. \ref{sec:model}.
The subband polarization and the maximization of the overlap via the variational pseudospin-polarization angle are discussed in Sec. 
\ref{sec:pol}. Section \ref{sec:results} is devoted to a discussion of our main results, and we present our conclusions in 
Sec. \ref{sec:concl}.

\section{Model}
\label{sec:model}

When restricted to the lowest LL, the leading energy scale for the electronic degrees of freedom is
set by the Coulomb interaction between electrons separated by the distance $r=|\rv_1-\rv_2|$ in the
plane,
\begin{align}\label{eq:potential}
V^{\sigma_1\dots \sigma_4}(r)& 
\nonumber \\
=\frac{e^2}{4\pi\epsilon} & \int_0^w \dr z \int_0^w \dr z' \; \frac{\varphi_{\sigma_1}^*(z)\varphi_{\sigma_2}^*(z') \varphi_{\sigma_3}(z)\varphi_{\sigma_4}(z')}{\sqrt{r^2+(z-z')^2}},
\end{align}
where $\varphi_{\sigma}(z)$ denotes the wave function in the $z$-direction associated with the first 
($\sigma=\downarrow$) and the second ($\sigma=\uparrow$) subband of the confinement potential. The 
wave functions of the two lowest subbands are linear combinations of Airy 
functions of the first and the second kind. A second
energy scale arises from the subband gap $\Delta(w,V)=E_2-E_1$, 
which is itself a function of the parameters
$w$ and $V$, 
and that may be obtained by solving the Schr\"odinger 
equation corresponding to the quantum mechanical problem
of a particle in a the potential shown in Fig. \ref{fig:TriangularEigen}. The subband gap 
plays the role of a Zeeman effect for the pseudospin. Notice furthermore that the third subband, which we neglect here, is found
at an energy $E_3-E_1\sim 3\Delta(w,V)$. In second quantized form, our model Hamiltonian
reads
\begin{align}\label{eq:general_hamiltonian}
 H= \frac{1}{2} \sum_{\{m_i\}}\sum_{\{\sigma_i\}} V_{m_1\dots m_4}^{\sigma_1\dots\sigma_4}(w,V)\; c^{\dagger}_{m_1\sigma_1} c^{\dagger}_{m_2\sigma_2} c_{m_4\sigma_4} c_{m_3\sigma_3} \nonumber \\-\frac{\Delta(w,V)}{2} \sum_m  (c^{\dagger}_{m\uparrow}c_{m\uparrow}-c^{\dagger}_{m\downarrow}c_{m\downarrow}),
\end{align}
where the operators $c_{m,\sigma}^{(\dagger)}$ annihilate (create) an electron in the lowest-LL
state $|n=0,m\rangle$ in the subband $\sigma$, where $m$ is the quantum number associated with the 
angular momentum. Here, we do not account for the physical spin degree of freedom and consider \textit{spinless}
fermions. A correct implementation of the physical spin, in addition to the subband pseudospin, would require a 
four-component treatment similarly to the case of graphene.\cite{GoerbigRegnault,PapicGraphene} This would 
substantially reduce the number of particles that can be diagonalized. Instead, we
rely on the fact that exchange effects generally favor spin-polarized states for a (spin) SU(2)-symmetric
interaction that are further stabilized by the Zeeman effect.
The matrix elements $V_{m_1\dots m_4}^{\sigma_1\dots\sigma_4}(w,V)$ correspond to the
two-body interaction (\ref{eq:potential}) when projected to the lowest LL. A large number of
non-zero matrix elements needs to be taken into account, in contrast to SU($2$)-symmetric spin 
systems or bilayer quantum Hall systems, where $\sigma_1=\sigma_3$ and $\sigma_2=\sigma_4$. Notice,
however, that the latter (bilayer) approximation was made in Ref. \onlinecite{Scarola}, where the FQHE
at half-filling was investigated via Monte-Carlo calculations.

\begin{figure}[htb]
\centering
\caption*{} 
\addtocounter{figure}{1} 
\subfloat[$\langle S^2 \rangle$\label{sub:S2_l_0}]{\includegraphics[scale=1]{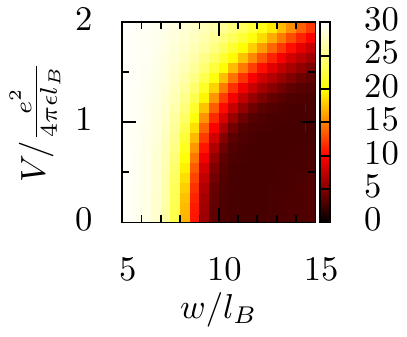}} \quad
\subfloat[$\langle S_z \rangle$ \label{sub:Sz_l_0}]{\includegraphics[scale=1]{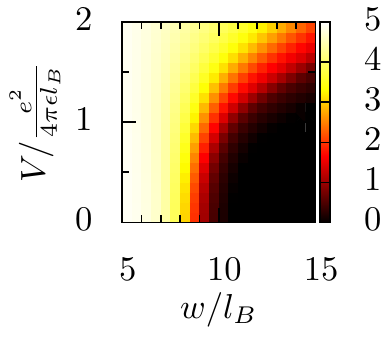}}
\addtocounter{figure}{-1} 
\caption{(Color online) 
Norm and $z$-component of pseudospin for $N=10$ and $N_B=17$. \label{fig:spins_l_0}}
\end{figure}

Within exact-diagonalization studies in the spherical geometry, we have investigated the ground-state
properties at half-filling in the lowest LL for $N=6$, $8$, and $10$ particles. The number of flux
quanta threading the sphere is $N_B=2N+\delta$, where $\delta$ is a state-dependent shift that is 
relevant in finite-size studies on the sphere. Here, we are interested in the Halperin
(331) state \cite{Halperin}
\begin{align}\label{eq:Halperin}
\Psi_{(331)}(\{z_i^{\uparrow}\}&,\{z_i^{\downarrow}\})=
 \prod_{i<j\leq N/2} (z_i^{\uparrow} - z_j^{\uparrow})^{3}  
 \nonumber \\ &\times \prod_{i<j\leq N/2} (z_i^{\downarrow} - z_j^{\downarrow})^{3} \prod_{i, j\leq N/2}  (z_i^{\uparrow} - z_j^{\downarrow}).
\end{align}
which has a shift $\delta=-3$.\cite{DeGail}
In Eq. (\ref{eq:Halperin}), the complex positions $z_j^{\sigma}$ are
those of particles in the pseudospin state $\sigma$, and we have omitted the ubiquitous Gaussian 
factor. 
The Hilbert space for $10$ particles (and $N_B=17$) is of dimension $1.2\times 10^{7}$
and the largest system size accessible numerically.
In the remainder of the paper, we present 
our results only for 10 particles since the results for $N=6$ and 8 are qualitatively the same.
We have concentrated our investigation on the experimentally relevant parameter range $w=5...15 l_B$ 
and $V=0...2e^2/\epsilon l_B$, where the ground state is always found in the sector with a 
total angular momentum $L_z=0$ that indicates a homogeneous charge distribution, necessary for 
possible FQHE states. Here and in the following sections, we measure lengths in units of the magnetic
length $l_B\simeq 25.8$ nm$/\sqrt{B{\rm [T]}}$ and energies in units of the Coulomb energy 
$e^2/\epsilon l_B$, where $\epsilon$ is the dielectric constant of the host semiconductor material.

\section{Subband polarization and rotation of the subband pseudospin}
\label{sec:pol}

Figure \ref{fig:spins_l_0} shows the average pseudospin polarization  
$\langle S^2\rangle$ and its $z$-component $\langle S_z\rangle$ in the ground state. 
These results are a first indication 
for a transition from one- to two-component states. Indeed, for narrow quantum wells with small 
values of $w$, the subband gap is too large to allow for a population of the higher subband 
$\sigma=\uparrow$, and the system is fully polarized in the $z$-direction -- one retrieves the 2D limit where only
the lowest subband is relevant.
That is also the case in larger wells with a high bias $V$ that effectively
reduces the well width by creating a triangular confinement potential. This needs to be contrasted to 
$w\sim 7l_B$, where one notices a reduction of the pseudospin polarization (both for 
$\langle S^2\rangle$ and $\langle S_z\rangle$) that indicates an increase in the population of
the upper subband. In the black region of Fig. \ref{fig:spins_l_0}, the vanishing
values for both quantities are a strong indication for a transition to a pseudospin-singlet state
with equal population of both subbands. Notice that in the full parameter range the polarization $\langle S_z\rangle$ 
never becomes negative such that there is no population inversion between the lowest two subbands. The absence of
such population inversion corroborates the validity of the model assumption to restrict the electron dynamics to
the lowest two subbands.

\begin{figure}[!ht]
\includegraphics[scale=0.8]{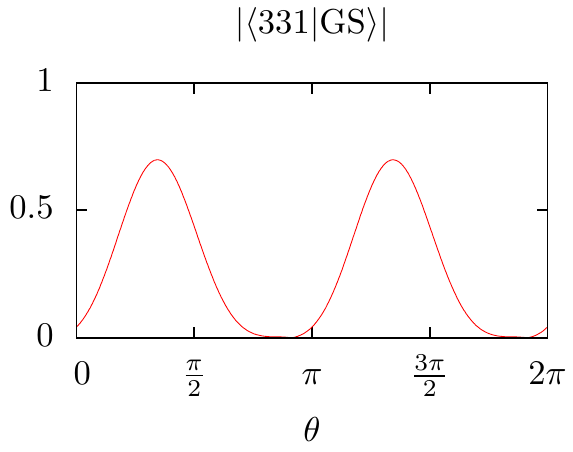}
\caption{(Color online)
Modulus of the overlap of (331) with the ground state in the lowest LL versus spin-rotation angle of (331),  here $w=10$, $V=1$ and the number of electrons is $N=10$.\label{fig:Ovl331VsTheta}}
\end{figure}

In order to interpret the results for the polarization, one needs to realize that, even if
the Halperin (331) state has a polarization $S_z=0$ in the $z$-direction, it is not a singlet since
it is not an eigenstate of the pseudospin $S^2$. As a consequence, the (331) state is not invariant under rotations of
the pseudospin, and one finds indeed that it possesses an average polarization 
$\langle S^2\rangle\neq 0$. Under rotation in the appropriate direction, the (331) state may thus be encountered
for intermediate pseudospin polarizations, that is for $\langle S_z\rangle\neq 0$ such as
in the red region of Fig. \ref{fig:spins_l_0}
that separates the completely polarized system (white) at low $w$ and high $V$ from the (black) singlet
region at large values of $w$ and low bias $V$. This rotation towards an appropriate frame of reference may be
implemented for the trial wave functions with the help of a variational parameter $\theta$, 
that allows one to maximize the overlap of the ground state with
the (331) state and other possible candidate wave functions. The appropriate transformation on the electron coordinates reads
\begin{equation}\label{eq:spinors_rotation}
\begin{pmatrix}
z_i^{\uparrow}  \\ 
z_i^{\downarrow}
\end{pmatrix} \longmapsto
\begin{pmatrix}
\cos\frac{\theta}{2} & -\sin\frac{\theta}{2} \\
\sin\frac{\theta}{2} & \cos\frac{\theta}{2}
\end{pmatrix} 
\begin{pmatrix}
 z_i^{\uparrow} \\
 z_i^{\downarrow}
\end{pmatrix}.
\end{equation} 

The strength of this variational approach is illustrated in Fig.
\ref{fig:Ovl331VsTheta}, 
where we have depicted the evolution of the overlap of the (331) state with the ground
state as a function of $\theta$ for a set of parameters ($w=10$ and $V=1$) that corresponds
precisely to an intermediate pseudospin polarization. Whereas the overlap almost 
vanishes for $\theta=0$
that is in the absence of a rotation and remains small in the $\pi/2$-rotated frame used in
the study of the symmetric quantum well,\cite{Papic} it exceeds $70\%$ for an optimal angle of
$\theta_{\rm opt}\simeq 1.08$. The $\pi$ periodicity in the angle-dependence of the overlap reflects
the fact that the $z$-component of the pseudospin polarization in the rotated frame vanishes for
both $\theta_{\rm opt}$ and $-\theta_{\rm opt}$, in contrast to maximally polarized states that
show a $2\pi$ periodicity (results not shown). Notice finally that the rotation of the (331) allows
us to obtain a state with partial subband polarization ($\langle S_z\rangle \neq 0$)
although one maintains $\langle S_z\rangle =0$ in the rotated frame.

\begin{figure}[!ht]
\caption*{} 
\addtocounter{figure}{1} 
\subfloat[$|\langle (331)_{\mathrm{opt}}|\mathrm{GS} \rangle|$ \label{sub:OvlOptimized}]{\includegraphics[scale=1]{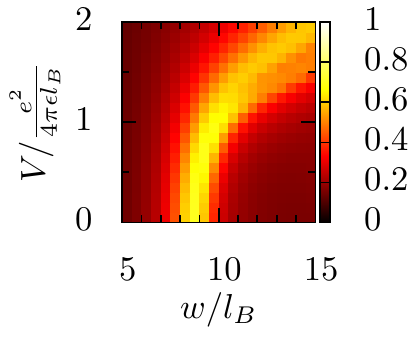}} \;
\subfloat[$\theta_{\mathrm{opt}}-\theta_{\mathrm{SQW}}$\label{sub:ThetaMax331}]{\includegraphics[scale=1]{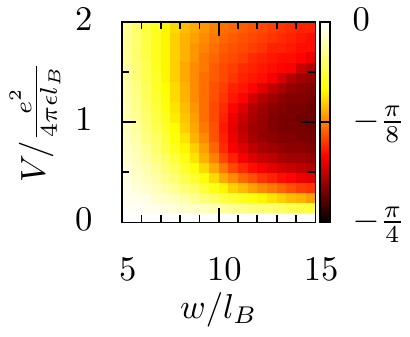}} \\
\subfloat[$|\langle (331)_{\mathrm{pol}}|\mathrm{GS} \rangle|$ \label{sub:OvlPol}]{\includegraphics[scale=1]{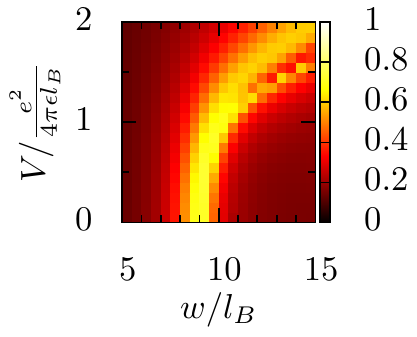}} \;
\subfloat[$\theta_{\mathrm{pol}}$ \label{sub:Polarization}]{\includegraphics[scale=1]{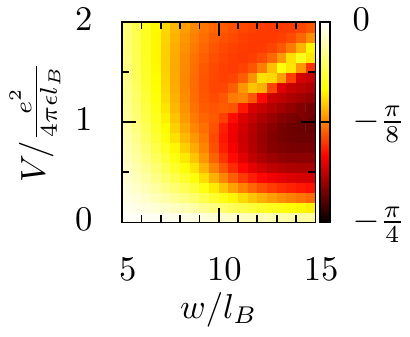}}
\addtocounter{figure}{-1} 
\caption{(Color online) (a) Optimized overlap between the ground and the (331) Halperin state for 
for $N=10$, $N_B=17$ in the lowest LL. (b) Angle that optimizes the above overlap with respect
to the angle $\theta_{\mathrm{SQW}}=\pi/2$, which is the optimal angle in the symmetric square
well.\cite{Papic} (c) Overlap between the ground and the (331) state for the mean
polarization angle~(\ref{eq:polarization}), which is depicted in (d).\label{fig:opt_and_pol}}
\end{figure}

\section{Results and discussion}
\label{sec:results}

\subsection{Parameter regions of maximal overlap}
\label{sec:maxOv}

The central result of our paper is shown in Fig. \ref{fig:opt_and_pol}, where we present the
optimal overlap [Fig. \ref{fig:opt_and_pol}(a)] between the ground state and the (331) Halperin state
at the rotation angle $\theta_{\rm opt}$ that maximizes the overlap within our
variational procedure. We have chosen a magnetic field of $B=14.2$ T, such as to make
a comparison with recent experimental results.\cite{Shabani2} To corroborate our findings, we present 
results for a different value of the magnetic field ($B=7$ T)in Sec. \ref{sec:diffB}.
The value of this angle is represented in Fig. \ref{fig:opt_and_pol}(b), with
respect to the reference angle $\theta_{\rm SQW}=\pi/2$ of the symmetric square well with
no bias investigated in Ref. \onlinecite{Papic}. 
One notices that the maximal overlap, which is in the $70\%$ range, is
indeed situated in the arc-shaped region of intermediate pseudospin polarization depicted in Fig.
\ref{fig:spins_l_0}. Concerning the black regions in Fig. \ref{fig:opt_and_pol}, where the 
overlap with the (331) state vanishes, we have found that the singlet state at large $w$ and
small $V$ has a reasonably high overlap with the compressible 
Haldane-Rezayi state,\cite{HaldaneRezayi} whereas in the small-$w$/large-$V$ limit
the Pfaffian \cite{MooreRead} has the largest overlap (results not shown). However, we stress that in
this limit of the lowest LL the (incompressible) Pfaffian is not the true ground state, but rather
the (compressible) composite-fermion Fermi liquid. Even if a direct comparison between these states is delicate in 
the spherical geometry, because the states do not occur at the same shift, we have checked that the
ground state above the arc-shaped region in Fig. \ref{fig:opt_and_pol}(a) (at $N=10$ and $N_B=18$, i.e. for 
a shift $\delta=-2$) is adiabatically connected to the composite-fermion Fermi liquid at $w=V=0$.
The incompressible Halperin state is thus surrounded by compressible states, and the FQHE can therefore 
be induced by increasing the bias for the appropriate well width.

\subsection{Optimal pseudospin angle}
\label{sec:optAng}

In order to obtain a better insight into the meaning of the angle $\theta(w,V)$
that optimizes the overlap, we consider again the pseudospin polarization of the ground state. 
Its average direction is characterized by the angle 
\begin{equation}\label{eq:polarization}
\theta_{\mathrm{pol}}=\arctan\left (\frac{\langle S_x \rangle}{\langle S_z \rangle}\right ),
\end{equation}
which only depends on the $x$- and the $z$-component of the polarization because the 
$y$-component vanishes, $\langle S_y\rangle =0$. This is a consequence of the considered
rotation (\ref{eq:spinors_rotation}), which only concerns the azimuthal angle $\theta$, and
we have tested numerically that $\langle S^2\rangle=\langle S_z^2\rangle + \langle S_y^2\rangle$
remains invariant under this transformation.
The polarization angle $\theta_{\rm pol}$ is shown
in Fig. \ref{fig:opt_and_pol}(d) as a function of the parameters $w$ and $V$, and one notices
an excellent agreement with the angle that optimizes the overlap between the ground and the 
(331) state in Fig. \ref{fig:opt_and_pol}(b). Furthermore, we have depicted in Fig. 
\ref{fig:opt_and_pol}(c) the same overlap for the polarization angle $\theta_{\rm pol}$ instead of
the optimal angle, and one finds that the difference between the overlaps is incremental in 
the region of interest. Our results therefore indicate that, in order to investigate the
(331) Halperin state in a BQW, one needs to use a rotated basis that is
determined by the polarization angle of the sublevel pseudospin. 

\subsection{Energy gaps}
\label{sec:gaps}

We now turn to the study of neutral energy gaps that is the minimal gap between the $L=0$ ground state and the first excited state (generally
at $L\neq 0$) at fixed particle and flux number. In the case of Laughlin states,\cite{Laughlin} this gap corresponds to the so-called 
magneto-roton minimum.\cite{GMP} Although this is not the activation gap obtained in magneto-transport measurements, 
the neutral gaps provide a measure of the ground-state stability.
The neutral energy gaps are presented in Fig.~\ref{sub:Gaps}, in comparison with the overlap of the (331) state with the ground state
[Fig.~\ref{sub:Ovl_n_8}, same as Fig.~\ref{sub:OvlOptimized}], for the optimal pseudospin-rotation angle $\theta$ of the former. 
As in the previous discussion, we show results for $N=10$ particles and $N_B=17$ flux quanta.

\begin{figure}[!ht]
\caption*{} 
\addtocounter{figure}{1} 
\subfloat[Gap in \% of $\tfrac{e^2}{4\pi\epsilon l_B}$ \label{sub:Gaps}]{\includegraphics[scale=1]{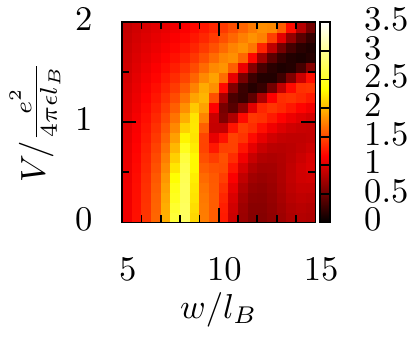}}
\subfloat[$|\langle (331)_{\mathrm{opt}}|\mathrm{GS} \rangle|$\label{sub:Ovl_n_8}]{\includegraphics[scale=1]{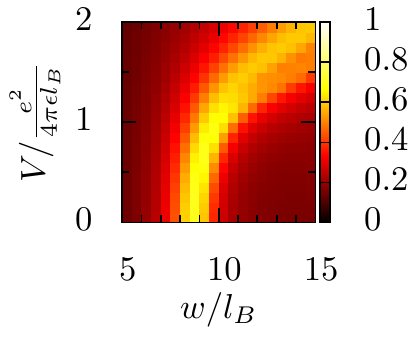}} \\
\addtocounter{figure}{-1} 
\caption{(Color online)
(a) Neutral energy gap [in \% of the Coulomb energy $e^2/(4\pi\epsilon l_B)$] and (b) optimized overlap of (331) with the ground state for $N=10$ and $N_B=17$.\label{fig:gaps}}
\end{figure}

The gaps are largest ($\sim 0.02...0.035 e^2/4\pi\epsilon l_B$)
in the upper-left part of the region of highest overlap (75 \%). Our results for the neutral gaps therefore 
corroborate the overall picture that the (331) state is stable in the parameter range where one finds the largest overlap with the ground state
obtained within exact diagonalization. 
Interestingly, one obtains a rather abrupt decrease of the gap below the arc-shaped region of maximal gap, while it remains substantial 
($\sim 0.01 e^2/4\pi\epsilon l_B$) in the region above the gap, in which one expects a polarized state. Whereas the origin of this decrease remains
to be understood, as well as its relative increase again at lower values of $V$ (and larger values of $w$), one notices that it is anticipated by
the overlaps (Fig. \ref{sub:Ovl_n_8}) where one finds a qualitatively similar fine structure in the arc-shaped region of the parameter space.

\subsection{Results for a different value of the magnetic field}
\label{sec:diffB}

In addition to a magnetic field $B=14.2$ T, a value used in the previous subsections, we present here results
for a different value $B=7$ T in order to test the global picture presented above. Figure \ref{fig:opt_and_pol_B_7} represents the 
same quantities for this value (for $N=10$ particles and $N_B=17$ flux quanta) as Fig. \ref{fig:opt_and_pol} for $B=14.2$ T.

\begin{figure}[!ht]
\caption*{} 
\addtocounter{figure}{1} 
\subfloat[$|\langle (331)_{\mathrm{opt}}|\mathrm{GS} \rangle|$ \label{sub:OvlOptimized_B_7}]{\includegraphics[scale=1]{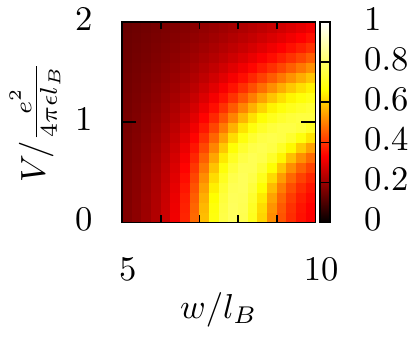}} \;
\subfloat[$\theta_{\mathrm{opt}}-\theta_{\mathrm{SQW}}$\label{sub:ThetaMax331_B_7}]{\includegraphics[scale=1]{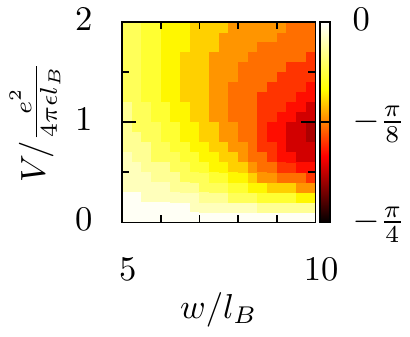}} \\
\subfloat[$|\langle (331)_{\mathrm{pol}}|\mathrm{GS} \rangle|$ \label{sub:OvlPol_B_7}]{\includegraphics[scale=1]{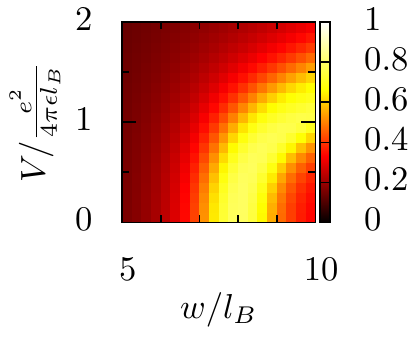}} \;
\subfloat[$\theta_{\mathrm{pol}}$ \label{sub:Polarization_B_7}]{\includegraphics[scale=1]{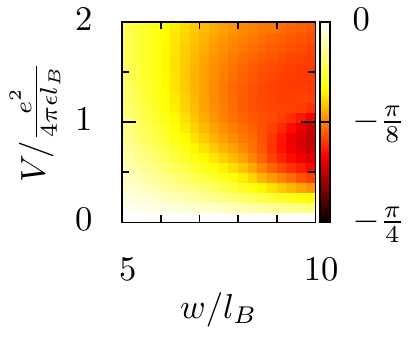}}
\addtocounter{figure}{-1} 
\caption{(Color online) Same quantities as Fig.~\ref{fig:opt_and_pol}, for a different value of the magnetic field, $B=7$ T.\label{fig:opt_and_pol_B_7}}
\end{figure}

One notices the similarity between the results for
$B=14.2$ T and $B=7$ T, are similar. 
For both values of the magnetic field the ground state of the system has a large overlap with the (331) state in an arc-shaped region of the 
parameter space. Quantitatively, two differences are worth pointing out.  
First, the region of high overlap mentioned above is shifted to slightly lower values of the width for $B=7$ T as compared to 
$B=14.2$ T. This shift as a function of the magnetic field may be understood qualitatively from simple scaling arguments. Consider the 
$V=0$ limit in which case the (331) state is stabilized when the gain in correlation energy $\sim (e^2/\epsilon l_B)\times(w/l_B)$ from 
filling the higher subband roughly cancels the subband gap $\Delta=\Delta(w,V=0)$,\cite{Papic} i.e. in our units
\begin{equation}
 \frac{\Delta}{e^2/\epsilon l_B}= c \frac{a_B}{l_B}\times\left(\frac{l_B}{w}\right)^2\sim \frac{w}{l_B},
\end{equation}
where $c$ is a numerical prefactor and $a_B=\hbar^2\epsilon/me^2$ is the effective Bohr radius in terms of the electronic band mass $m$. This 
leads to a $B$-field scaling
\begin{equation}
 \frac{w}{l_B} \sim \left(\frac{a_B}{l_B}\right)^{1/3}\propto B^{1/6}
\end{equation}
for the characteristic parameter at which the (331) state is stabilized. In the present case (Fig. \ref{fig:opt_and_pol_B_7}), 
the magnetic field is half of that discussed previously in Fig. \ref{fig:opt_and_pol}, such that one expects for the characteristic
width parameter $(w/l_B)[B=7~\rm{T}]\simeq 0.9(w/l_B)[B=14~\rm{T}]$, in agreement with our numerical findings.

As for the second quantitative difference between the two values of the magnetic field, one notices that 
the overlap of the ground state and the (331) state is slightly enhanced for $B=7$ T. It is on the order of $80\%$ in the high 
overlap region whereas it is $\sim 70\%$ for $B=14.2$ T.

\subsection{Comparison with experiments}
\label{sec:exp}

Notice finally that our variational approach provides an excellent understanding of recent 
magneto-transport measurements by Shabani \textit{et al.}, who investigated quantum-Hall states
in quantum wells with a potential bias applied between the sides.\cite{Shabani2,Shabani3}
In the case of the 55-nm wide sample, which we discuss as a representative example here,
half-filling of the lowest LL is encountered at a magnetic field of $14.2$ T, which is 
precisely the value we have chosen in our numerical studies. It corresponds to a ratio 
of $w/l_B\simeq 8$ between the well width and the magnetic length. A FQHE could be stabilized 
for values of the bias around $V=16.5$ meV that correspond to $V/(e^2/4\pi\epsilon l_B)\simeq 1$ 
in our units. Our overlap calculations in Fig. \ref{fig:opt_and_pol}
indicate a similar window for the observation of a FQHE slightly below $w\simeq 10 l_B$ that
is some $10-20\%$ larger than the experimental value. Apart from the finite size of the system
studied numerically, this slight discrepancy is likely to be due
to our BQW model that considers infinite potential barriers, whereas the potential barriers 
are finite ($\sim$ 200 meV) in a realistic situation such that the wave functions $\varphi_{\sigma}(z)$ can
penetrate into the barriers. The wave function in the $z$-direction is therefore slightly
larger than in the BQW model, where the width of the wave function coincides with the well
width $w$. To compensate this effect, one therefore needs to consider a larger effective width
within our model. 

\section{Conclusions}
\label{sec:concl}

In conclusion, we have presented numerical evidence that a FQHE in the form of a (331) Halperin
state can be stabilized in a wide quantum well with a potential bias in an intermediate range,
as observed experimentally. From a technical point of view, we have established a model (BQW) that 
allows one to obtain the wave functions, which are associated with the confinement potential.
They are the basic ingredient in the calculation of the effective interaction potential, which
is used in our exact-diagonalization studies on the sphere. 
This model may be useful in further studies of the FQHE in wide anisotropic quantum wells. 
Furthermore, we have shown that the wave-function overlap between the ground state and
trial wave functions, as the (331) state here, can be largely enhanced if the state is prepared
in a rotated frame that corresponds to the orientation of the sublevel-pseudospin polarization. 
We expect that this rotated frame is also essential in the understanding of other possible 
FQHE states in the BQW, beyond the (331) state.

We thank Gilles Montambaux for an optimized initial guess of BQW eigenergies.
Furthermore, we acknowledge fruitful discussions with Paul Soul\'e and Zlatko Papi\'c. N.R. was supported by 
a Keck grant.

\bibliography{TriangularFQHE}

\end{document}